\documentclass[12pt]{iopart}

\usepackage{amsfonts}

\usepackage{amscd,amsthm}

\usepackage{graphicx}

\usepackage{color,ulem}

\theoremstyle{plain} 

\newtheorem{theorem}{Theorem}[section]

\newtheorem{remark}[theorem]{Remark}

\begin{document}

\title{Collective observables in repeated experiments of population dynamics}

\author{T. Carletti}

\address{D\'epartement de math\'ematique, Facult\'es
  Universitaires Notre Dame de la 
  Paix, rempart de la Vierge 8, B 5000 Namur, Belgium}

\author{D. Fanelli}

\address{Dipartimento di Energetica and CSDC, Universit\`a di
  Firenze, and INFN, via S. Marta, 3, 50139 Firenze, Italy}

\begin{abstract}
We here discuss the  outcome of an hypothetic experiments of populations dynamics, where a set 
of independent realizations is made available. The importance of ensemble 
average is clarified with reference to the registered time evolution of key collective 
indicators. The problem is here tackled for the logistic case study. Theoretical 
prediction are compared to numerical simulations. 
\end{abstract}

\maketitle

\section{Introduction}
\label{sec:intro}

The problem of explaining the emergence of self-organized, macroscopic, patterns from a limited set of rules 
governing the mutual interaction of a large assembly of microscopic actors, is often faced in several domains of physics 
and biology. This challenging task defines the realm of complex systems, and calls for novel paradigms 
to efficiently intersect distinct expertise. 

Population dynamics has indeed attracted many scientists \cite{murray} and dedicated 
models were put forward to reproduce in silico the change in population over time as displayed in real ecosystems (including 
humans). Two opposite tendencies are in particular to be accomodated for. On the one hand, microscopic agents do reproduce themselves 
with a specific rate $r$, an effect which translates into a growth of the population size $P$. On the other, competition for the available 
resources (and death) yields a compression of the population. In a seminal work by Verhulst \cite{verhulst}, these 
ingredients were formalized in the differential equation:
\begin{equation}
  \label{eq:Ver}
  \frac{d P}{d t}= r P\left(1-\frac{P}{K}\right).
\end{equation}
$K$ is the so called carrying capacity and identifies the
maximum allowed population for a selected organism,
under specific environmental  
conditions. The above model predicts an early exponential growth, which is subsequently antagonized by the quadratic contribution, responsible 
for the asymptotic saturation. The adequacy of the Verhulst's model was repeatedly tested versus laboratory experiments:
Colonies of bacteria, yeast or other simple organic entities were grown, while monitoring the time evolution of the population amount. In some cases,
an excellent agreement \cite{Krebs,Stogatz} with the theory was reported, thus supporting the biological validity of Eq.~(\ref{eq:Ver}). 
Conversely, the match with the theory was definetely less satisfying for e.g. fruit flies, flour beetles and in general for other
organisms that rely on a more complex life cycle. For those latter, it is necessary to invoke a somehow richer modelling scenario 
which esplicitly includes age structures and time delayed effects of
overcrowding population \cite{Stogatz}. For a
more deailed account on these  issues the interested reader can refer to the review paper \cite{Krebs} and references therein. 

Clearly, initial conditions are crucial and need to be accurately determined. An error in assessing the 
initial population, might reflect in the estimates of the parameters $r$ and $K$, which are tuned so to adjust  
theoretical and experimental data. In general, the initial condition relative to one specific experimental 
realization could be seen as randomly extracted from a given distribution. This, somehow natural, viewpoint is elaborated 
in this paper and its implications for the analysis of the experiments thoroughly explored.

In particular we shall focus on the setting where $N$ independent population
communities are (sequentially or simultaneously) made to evolve.
The experiment here consists in measuring collective observables, as the average population and associated momenta of the 
ensemble distribution.  As anticipated, sensitivity to initial condition do play a crucial role and so need to be properly addressed when aiming at establishing a link with 
(averaged) ensemble measurements, or, equivalently, drawing reliable forecast. To this end, we will here develop two analytical approaches which enable us to reconstruct the sought distribution. The first,
to which section \ref{sec:momevo} is devoted, aims at obtaining a complete
description of the momenta, as e.g. the mean population amount. This is an observable of paramount importance,  
potentially accessible in real experiments. The second, discussed in section
\ref{sec:pdfevo},  
introduces a master equation which rules the evolution of the relevant
distribution. It should be remarked that this latter approach is a priori more general then the former, 
as the momenta can in principle be calculated on the basis of the recovered distribution. However,
computational difficulties are often to be faced which make the analysis
rather intricate. In this perspective the two proposed scenario are to be
regarded as highly complementary.

 In the following, for practical purposes, we shall assume each population to
 evolve as prescribed by a Verhulst type of equation. The  
 methods here developed are however not limited to this case study but can be
 straightforwardly generalized to settings were other, possibly more 
 complex, dynamical schemes are put forward.

\section{On the momenta evolution}
\label{sec:momevo}

Imagine to label with $x_i$ the population relative to the $i$-th
realization, belonging to the ensemble of $N$ independent
replica. As previosuly recalled, we assume each $x_i$ to obey a first order
differential equation of the logistic type,  namely:
\begin{equation}
  \label{eq:Ver1}
  \frac{d x_i}{d t}= x_i (1-x_i)\, ,
\end{equation}
that can be straightforwardly obtained from~(\ref{eq:Ver}) by setting
  $x=P/K$ and renaming the time $t\rightarrow r t$. The initial condition
  will be denoted by $x_i^0$.  

A natural question concerns the expected output of
an hypothetic set of experiments constrained as above. More 
concretely, can we describe the distribution of possible solutions, once the collection of initial data is entirely specified?

The $m$-th momentum associated to the discrete distribution
of $N$ repeated measurements acquired at time $t$ reads:
\begin{eqnarray}
  \label{eq:mmom2}
  <x^{m}>(t)=\frac{\left(x_1(t)\right)^m+\dots+\left(x_N(t)\right)^m}{N}\, ,
\end{eqnarray}

To reach our goal, we introduce the {\it time dependent moment
  generating function}, $G(\xi,t)$, 
\begin{equation}
  \label{eq:formF}
  G(\xi,t):=\sum_{m=1}^{\infty}\xi^m <x^m>(t)\, .
\end{equation}
This is a formal power series whose Taylor coefficients are the momenta
of the distribution that we are willing to reconstruct, task that can 
be accomplished using the following relation: 
\begin{equation}
  \label{eq:momen}
  <x^m>(t):=\frac{1}{m!}\frac{\partial^m G}{\partial
    \xi^m}\Big |_{\xi=0}\, . 
\end{equation}

By exploiting the evolution's law for each $x_i$, 
we shall here obtain a partial differential equation governing
  the behavior of $G$. Knowing $G$  
will eventually enables us to calculate any sought momentum via multiple
differentiation with respect to $\xi$ as stated in~(\ref{eq:momen}).

Deriving~(\ref{eq:mmom2}) and making use of Eq.~(\ref{eq:Ver1}) immediately
yields: 

\begin{eqnarray}
  \label{eq:xmom}
  \frac{d}{dt}<x^m>(t)&=&\frac{1}{N}\sum_{i=1}^N
  \frac{dx_i^m}{dt}=\frac{m}{N}\sum_{i=1}^N x_i^{m-1} \frac{dx_i}{dt}\nonumber
  \\
&=&\frac{m}{N}\sum_{i=1}^N x_i^{m-1}
x_i(1-x_i)=m\left(<x^m>-<x^{m+1}>\right)\, ,
\end{eqnarray}

On the other hand, 
by differentiating~(\ref{eq:formF}) with respect to
time, one obtains : 
\begin{equation}
  \label{eq:diff1}
  \frac{\partial G}{\partial t}=\sum_{m\geq 1}\xi^m\frac{d<
    x^m>}{dt}=\sum_{m\geq 1}m\xi^m\left(<x^m>-<x^{m+1}>\right)\, , 
\end{equation}
where used has been made of Eq.~(\ref{eq:xmom}). We can now re-order the terms so to express the right
hand side as a function of $G$~\footnote{Here 
 the following algebraic relations are being used:
\begin{equation}
  \label{eq:derx}
  \xi\partial_{\xi} G(\xi,t)=\xi\sum_{m\geq 1}m\xi^{m-1}<x^m> = \sum_{m\geq
    1}m\xi^{m}<x^m> \, , \nonumber
\end{equation}
and 
\begin{eqnarray}
  \label{eq:der2x}
  \xi\partial_{\xi} \frac{G(\xi,t)}{\xi}&=&\xi\partial_{\xi}\sum_{m\geq
    1}\xi^{m-1}<x^m> = 
  \xi\sum_{m\geq 1}(m-1)\xi^{m-2}<x^m>\nonumber \\ &=&\sum_{m\geq
    1}(m-1)\xi^{m-1}<x^m> \, \nonumber
\end{eqnarray}
Renaming the summation index, $m-1\rightarrow m$, one finally gets (note
the sum still begins with $m=1$):
\begin{equation}
  \label{eq:derxx}
  \xi\partial_{\xi} \frac{G(\xi,t)}{\xi}=\sum_{m\geq 1}m\xi^{m}<x^{m+1}>
  \,\nonumber.
\end{equation}}
and finally obtain the following non--homogeneous linear partial differential
equation:  
\begin{equation}
    \label{eq:forF}
  \partial_t G - (\xi-1)\partial_{\xi} G =\frac{G}{\xi}\, .
\end{equation}

Such an equation can be solved for $\xi$ close to zero (as in the end of the procedure 
we shall be interested in evaluating the derivatives at 
$\xi=0$, see Eq. (\ref{eq:momen})
) and for all positive $t$. To this end we shall specify the 
initial datum: 
\begin{equation}
  \label{eq:inidat}
  G(\xi,0)=\sum_{m\geq 1}\xi^m<x^m>(0)=\Phi(\xi)\, ,
\end{equation}
i.e. the initial momenta or their distribution.

Before turning to solve~(\ref{eq:forF}), we first simplify it by introducing
\begin{equation} 
  \label{eq:fF}
  G=e^g \quad {\rm namely}\quad g=\log G\, ,
\end{equation}
then for any derivative
\begin{equation}
  \partial_* G = G\partial_*g \, ,
\end{equation}
where $*=\xi$ or $*=t$, thus~(\ref{eq:forF}) is equivalent to
\begin{equation}
  \label{eq:forf}
  \partial_t g-(\xi-1)\partial_{\xi} g =\frac{1}{\xi}\, ,
\end{equation}
with the initial datum
\begin{equation}
  \label{eq:initdatf}
  g(\xi,0)=\phi(\xi)\equiv \log \Phi(\xi)\, .
\end{equation}

This latter equation can be solved using the {\it method of the
  characteristics}, here represented by:
\begin{equation}
  \label{eq:char}
  \frac{d\xi}{dt}=-(\xi-1)\, ,
\end{equation}
which are explicitly integrated to give:
\begin{equation}
  \label{eq:charsol}
  \xi(t)=1+(\xi(0)-1)e^{-t}\, ,
\end{equation}
where $\xi(0)$ denotes $\xi(t)$ at $t=0$. Then the function
$u(\xi(t),t)$ defined by: 
\begin{equation}
  \label{eq:funcu}
  u(\xi(t),t):=\phi(\xi(0))+\int_0^t\frac{1}{1+(\xi(0)-1)e^{-s}}\, ds\, , 
\end{equation}
is the solution of~(\ref{eq:forf}), restricted to the
characteristics. Observe that $u(\xi(0),0)=\phi(\xi(0))$, so~(\ref{eq:funcu})
solves also the initial value problem. 

Finally the solution of~(\ref{eq:initdatf}) is obtained from $u$ by reversing
the relation between $\xi(t)$ and $\xi(0)$, i.e. $\xi(0)=(\xi(t)-1)e^t+1$:
\begin{equation}
  \label{eq:soluf}
  g(\xi,t)=\phi\left((\xi-1)e^t+1\right)+\lambda(\xi,t)\, ,
\end{equation}
where $\lambda(\xi,t)$ is the value of the integral in the right hand side
of~(\ref{eq:funcu}). 

This integral can be straightforwardly computed as follows (use the
change of variable $z=e^{-s}$):
\begin{equation}
\label{eq:int1}
  \lambda = \int_0^t\frac{1}{1+(\xi(0)-1)e^{-s}}\, ds
  =\int_1^{e^{-t}}\frac{-dz}{z}\frac{1}{1+(\xi(0)-1)z}\, , 
\end{equation}
which implies 
\begin{eqnarray}
  \label{eq:int2}
  \lambda &=&
  -\int_1^{e^{-t}}dz\left(\frac{1}{z}-\frac{\xi(0)-1}{1+(\xi(0)-1)z}\right)= -\log 
  z+\log (1+(\xi(0)-1)z)\Big |_{1}^{e^{-t}}\nonumber \\ &=&t+\log
  (1+(\xi(0)-1)e^{-t})-\log \xi(0)\, . 
\end{eqnarray}

According to~(\ref{eq:soluf}) the solution $g$ is then
\begin{equation}
  \label{eq:soluf2}
  g(\xi,t)=\phi\left((\xi-1)e^t+1\right)+t+\log \xi -\log ((\xi-1)e^{t}+1)\, ,
\end{equation}
from which $G$ straightforwardly follows: 
\begin{equation}
  \label{eq:solFF}
  G(\xi,t)=\Phi\left((\xi-1)e^{t}+1\right)\frac{\xi e^t}{(\xi-1)e^{t}+1}\, .
\end{equation}

As anticipated, the function $G$ makes it possible to estimate any momentum  (\ref{eq:momen}). As an 
example, the mean value correspond to setting  $m=1$, reads: 
\begin{eqnarray}
  \label{eq:mom1f}
  <x>(t)&=&\partial_{\xi}G\Big |_{\xi=0}=\Big[
  \Phi^{\prime}\left(1+(\xi-1)e^{t}\right) e^{t}
  \frac{\xi e^t}{(\xi-1)e^{t}+1}\nonumber\\ &+&\Phi\left(1+(\xi-1)e^{t}\right)
  e^{t}\frac{(\xi-1)e^{t}+1-\xi
    e^{t}}{\left(1+(\xi-1)e^{t}\right)^2}\Big]\Big |_{\xi=0}\nonumber\\
  &=&\frac{e^{t}}{1-e^{t}}\Phi(1-e^{t})\, .   
\end{eqnarray}

In the following section we shall turn to considering a specific application 
and test the adequacy of the proposed scheme.

\section{Uniform distributed initial conditions}

In this section we will focus on a particular case study in the aim of
clarifying the potential interest 
of our findings. The inital data (i.e. initial population amount) are assumed to span uniformly a
bound interval $[a,b]$. No prior information is hence available which favours one specific choice, all accessible initial data being equally probably.
To fix the ideas we shall here set  $a=0$ and $b=1/2$. The
probability distribution $\psi(x)$ clearly reads~\footnote{We hereby assume to
  sample over a large collection of independent replica of the system under
  scrutiny 
(N is large). Under this hypothesis one can safetly adopt a continuous
approximation for the distribution of allowed initial data. Conversely, if the
number of  
realizations is small, finite size corrections need to be included.}:
\begin{equation}
  \psi(x)=  \cases{2 & if $x\in[0,1/2]$\\0 & otherwise}\, ,
\end{equation}
and cosequently the initial momenta are:
\begin{equation}
  \label{eq:mommexUD}
  <x^m>(0)=\int_0^1 \xi^m
  \psi(\xi) d\xi =\int_0^{1/2}2\xi^m\, d\xi = \frac{1}{m+1}\frac{1}{2^m}\, .
\end{equation}

Hence the function $\Phi$ as defined in~(\ref{eq:inidat}) takes the form:
\begin{equation}
  \label{eq:phiUD}
  \Phi(\xi)=\sum_{m\geq 1}\frac{1}{m+1}\frac{\xi^m}{2^m}\, .
\end{equation}
A straightforward algebraic manipulation allows us to
re-write~(\ref{eq:phiUD}) as follows:
\begin{equation}
  \sum_{m\geq 1}\frac{y^m}{m+1}=\frac{1}{y}\int_0^y \sum_{m\geq 1}z^m\,
  dz=\frac{1}{y}\int_0^y \frac{z}{1-z}\, dz = -1-\frac{1}{y}\log (1-y)\, ,
\end{equation}
thus
\begin{equation}
  \label{eq:phiUDfin}
  \Phi(\xi)=-1-\frac{2}{\xi}\log \left(1-\frac{\xi}{2}\right)\, .
\end{equation}

We can now compute the time dependend moment generating function, $G(\xi,t)$,
given by~(\ref{eq:solFF}) as:
\begin{equation}
  \label{eq:FUD}
  G(\xi,t)=\frac{\xi e^{t}}{(\xi -1)e^{t}+1}\left[-1-\frac{2}{(\xi -1)e^{t}+1}\log \left(
      1-\frac{(\xi -1)e^{t}+1}{2}\right)\right]\, ,
\end{equation}
and thus recalling~(\ref{eq:momen}) we get
\begin{eqnarray}
  \label{eq:ameda2t}
  <x>(t)&=&\frac{e^{t}}{e^{t}-1}-\frac{2e^{t}}{(e^{t}-1)^2}\log\left(\frac{e^{t}+1}{2}\right)\\ 
<x^2>(t)&=&\frac{e^{2 t}}{(e^{t}-1)^2}+\frac{4e^{2
    t}}{(e^{t}-1)^3}\log\left(\frac{e^{t}+1}{2}\right)+\frac{2e^{2t}}{(e^{t}-1)^2(e^{ t}+1)}\nonumber\, .
\end{eqnarray}

\begin{figure}[htbp]
\centering
\includegraphics[scale=0.4]{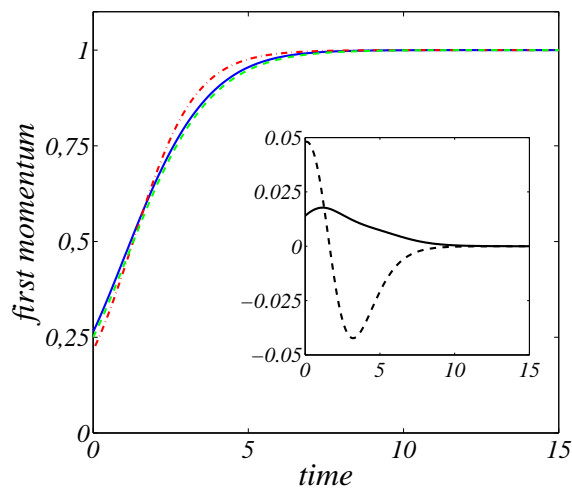}
\caption{Main panel: Time evolution of the first moment $<~x~>~(~t~)$. The (blue) solid line stands for direct simulations 
averaged over $N=100$ independent realizations. The (green) dashed line
represents the analytical solution  (\ref{eq:ameda2t}a).  
The (red) dot-dashed line is the solution of the logistic Eq.~(\ref{eq:Ver1}),
where the initial datum is being adjusted to the best fit value  
$\hat{x}^0 = 0.216$. Inset: the solid (resp. dashed) line represents the
difference between the analytical (resp. fitted) and numerical curves.}  
\label{fig:1mom}
\end{figure}

For large enough times, the distribution of the experiments' outputs
is in fact concentrated around the asymptotic value $1$ with an associated
variance (calculated from the above momenta) which decreases monotonously with time.
In Fig.~\ref{fig:1mom} direct numerical simulations are compared to the analytical solution~(\ref{eq:ameda2t}a), returning a 
good agreement. A naive approach would suggest interpolating the averaged numerical profile with a solution of the 
logistic model whose initial datum $\hat{x}^0$ acts as a free
  parameter to be adjusted to its best fitted value. As testified by
visual 
inspection of Fig.~\ref{fig:1mom}  this procedure yields a significant
discrepancy, which could be possibly misinterpreted as a failure of the
underlying   
logistic evolution law. For this reason, and to avoid drawing erroneous conclusions when ensemble averages are computed,  
attention has to be payed on the role of initial conditions. 

\begin{remark}[Best parameters estimates]
In the preceding discussion the role of initial condition 
was elucidated. In a more general setting one might imagine $r$, the logistic parameter, 
to be an unknown entry to the model (see Eq.~(\ref{eq:Ver1})). One could therefore imagine to
proceed with a fitting strategy which adjusts both  
$\hat{x}^0$ and  $r$ so to match the (averaged) data. 
Alternatively, and provided the distribution of initial conditions is assigned (here assumed uniform), 
one could involve the explicit solution~(\ref{eq:ameda2t}a)
where time is scaled back to ist original value: 

\begin{equation}
\label{theory_r} 
  <x>(t)=\frac{e^{rt}}{e^{rt}-1}-\frac{2e^{rt}}{(e^{rt}-1)^2}\log\left(\frac{e^{rt}+1}{2}\right)\, .
\end{equation}
and let the solely parameter $r$ to run freely so to search for the optimal agreement with the data.
As an example, we perfomed $N=100$ repetead numerical simulations of the
logistic model with parameter $r=1.5$ and intial data uniformly
distributed in $[0,1/2]$. Using the straightforward solution of the logistic 
equation where $\hat{x}^0$ and  $r$ are adjusted, returns $r=1.2123$. 
The analysis based on (\ref{theory_r}) leads to 
$r=1.5662$, which is definitely closer to the true value.
\end{remark}

\begin{remark}[On the case of a normal distribution]
The above discussion is rather general and clearly extends 
beyond the uniform distribution case study. The analysis can be in fact adapted to other
settings, provided the distribution of initially allowed population amount 
is known. We shall here briefly discuss the rather interesting case where a
normal distribution is to be considered. Let us assume that
$x^0_{i}$ are random  normally distributed values with 
mean $1/4$ and standard deviation $\sigma^2$, one can compute all the
intial momenta $<x^m>(0)$ as:
\begin{equation}
  \label{eq:mommex}
  <x^m>(0)=\int \xi^m
  \frac{1}{\sigma^2\sqrt{2\pi}}e^{-\frac{1}{2}\left(\frac{\xi-1/4}{\sigma}\right)^2}\,  d\xi \, .
\end{equation}
Assuming $\sigma^k$, $k\geq 3$ to be negligible with
respect to $\sigma^2$, 
the function $\Phi(\xi)$ specifying the initial datum
  in Eq.~(\ref{eq:inidat}) reads:
\begin{equation}
  \label{eq:funphi}
  \Phi(\xi)=\sum_{m\geq 1}<x^m>\xi^m =
  \frac{\xi}{4}+\frac{\xi^2}{4^2}+\sigma^2\xi^2+\sum_{m\geq
    3}\left[\left(\frac{\xi}{4}\right)^m+\frac{m(m-1)}{2}\left(\frac{\xi}{4}\right)^{m-2}\sigma^2\xi^2\right]\, .
\end{equation}
Collecting together the terms $(\xi/4)^m$ for $m\geq 1$ we obtain:
\begin{equation}
  \sum_{m\geq 1}\left(\frac{\xi}{4}\right)^{m}=\frac{\xi}{4-\xi}\, ,
\end{equation}
while the remaining terms read:
\begin{equation}
  \sigma^2\xi^2+\sum_{m\geq
    3}\frac{m(m-1)}{2}\left(\frac{\xi}{4}\right)^{m}4^2\sigma^2=\sum_{m\geq
    2}\frac{m(m-1)}{2}\left(\frac{\xi}{4}\right)^{m}4^2\sigma^2\, .
\end{equation}
It is then easy to verify that their contributution to the required $\Phi$
  funcion results in 
\begin{equation}
  \label{eq:fprphifin}
  \Phi(\xi)=\frac{\xi}{4-\xi}+\frac{(4\sigma)^2}{2}\frac{2(\xi/4)^2}{(1-\xi/4)^3}=\frac{\xi}{4-\xi}+\frac{4^3\xi^2\sigma^2}{(4-\xi)^3}\, .
\end{equation}

To proceed further we again calculate the derivatives of $G$
(defined through the function $\Phi$), evaluate them at $\xi=0$, and
eventually  get the evolution of $<x^m>$ in time, for all $m\geq 1$.
\end{remark}

\section{Monitoring the time evolution of the probability distribution
  function of expected measurements} 
\label{sec:pdfevo}

As opposed to the above procedure, one may focus on the distribution function of expected outputs, rather then computing its momenta.  
The starting point of the analysis relies on a generalized version of the celebrated Liouville theorem. This latter asserts 
that the phase-space distribution function $f$ is constant along the trajectory of the system. For a non Hamiltonian system this condition results 
in the following equation (for convenience derived in the Appendix
~\ref{sec:app}) for the evolution of the 
probability density function under the action of a generic ordinary differential equation, here represented by the vector field $\vec{X}$: 

\begin{equation}
  \label{eq:pdfevolv}
  \frac{\partial f}{\partial t}+\nabla f \cdot \vec{X}+f {\mathit
    div} \vec{X} = 0\, ,
\end{equation}
where ${\mathit div} \vec{X}=\sum\partial X_i/\partial x_i$. 

For the case under inspection the $1$--dim vector field reads  $\vec{X}(x) = x(1-x)$ and hence
${\mathit div} \vec{X} = (1-2x)$. Thus, introducing $F=\log f$
Eq.~(\ref{eq:pdfevolv}) can be cast in the form: 
\begin{equation}
  \label{eq:alphapdfeq}
  \frac{\partial F}{\partial t}+  x (1-x)\frac{\partial
    F}{\partial x}+(1-2x)=0\, .
\end{equation}
 
\begin{figure}[htbp]
\centering
\includegraphics[scale=0.5]{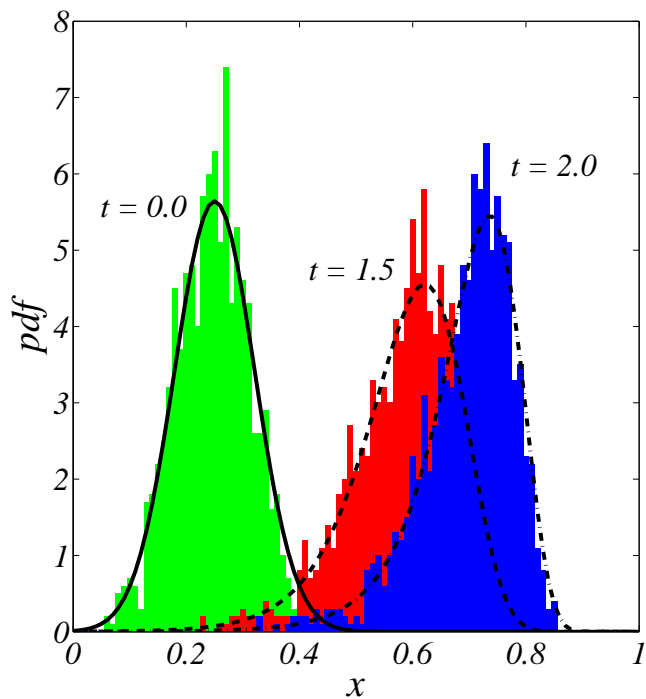}
\caption{
Time evolution of the probability distribution function. Histograms refers to numerical simulation and are calculated at different time:
 $t=0$ (green online),  $t=1.5$ (red, online),  $t=2.0$ (blue online). The lines represent the 
corresponding analytical solution}  
\label{fig:pdfevolvnorm}
\end{figure}

To solve this equation we use once again the methods of characteristics, which
are now solutions of $\dot{x}=x(1-x)$, namely:
\begin{equation}
  \label{eq:charact}
  x(t) = \frac{x(0)e^{t}}{1-x(0)+x(0)e^{t}}\, ,
\end{equation}
The solution of~(\ref{eq:alphapdfeq}) is hence:
\begin{equation}
  \label{eq:solF}
  F(x,t)=F_0(x(0))-\int_0^t (1-2x(s))\, ds \, ,
\end{equation}
where $F_0=\log \psi$ is related to the probability distribution function at $t=0$ and must
be evaluated at $x(0)$, seen as a function of $x(t)$. The integral can be
computed as follows:
\begin{equation}
  \label{eq:integ}
  \int_0^t (1-2x(s))\, ds=\int_0^t
  \left(1-2\frac{x(0)e^{s}}{1-x(0)+x(0)e^{s}}\right)\, ds=  t +2\log \left(
    1-x(0)+x(0)e^{t}\right)\, . 
\end{equation}
Such an expression has to be introduced into~(\ref{eq:solF}) once we explicit
$x(0)$ for $x(t)=x$ as:
\begin{equation}
  \label{eq:alpha0alphat}
  x(0) = \frac{x e^{- t}}{1-x+x e^{-t}}\, .
\end{equation}
Hence:
\begin{equation}
  \label{eq:Ffinal}
  F(x,t)=F_0\left(\frac{x e^{-t}}{1-x+x e^{-t}}\right)- t-2 \log  \left( 1-x+x
    e^{-t}\right)\, ,   
\end{equation}
and finally back to the original $f$:
\begin{equation}
\label{eq:ffinal}
  f(x,t)=\psi\left(\frac{x e^{- t}}{1-x+x e^{-t}}\right)\frac{e^{-t}}{\left(
      1-x+x e^{-t}\right)^2}\, , 
\end{equation}
which stands for the probability density function which describes for all $t$
the expected distribution of $x$'s. In Fig.~\ref{fig:pdfevolvnorm} we compare
the analytical solutions~(\ref{eq:ffinal}) with the numerical simulation of
the logistic model~(\ref{eq:Ver1}) under the assumption of $N=1000$ initial
data normally distributed with mean $1/2$ and variance $0.005$.

Notice that having calculated the distribution $f$ will enable in turn, at
least in principle, to  
to calculate all the associated momenta. 

\section{Conclusion}

Forecasting the time evolution of a system which obeys to a specifc governing differential 
equation and is initialized as follows a specific probability distribution, constitutes a central 
problem in several domains of applications. Assume for instance a set of independent measurements to return an
ensemble average which is to be characterized according to a prescribed model. Biased conclusion  
might result from straightforward fitting strategies which do not correctly weight the allowed distribution of initial condition. 

In this paper we address this problem
by providing an exact formula for the time evolution of momenta and probability distribution function of expected measurements,
which is to be invoked  for a repeaded set of indipendent experiments. Though general, the method is here discussed with 
reference to a simple, demonstrative problem of population 
dynamics.

\section{Acknowledgments}
We wish to thank M. Villarini for several discussion and, in particular, for 
suggesting Eq.~(\ref{eq:therelat}).

\appendix
\section{The generalized Liouville theorem}
\label{sec:app}
Let $\vec{X}(x)$ be a vector field to which
we 
associate the ordinary differential equation: 
  \begin{equation}
    \label{eq:ode1}
    \dot{x} = \vec{X}(x)\quad \forall x\in \Omega\, ,
  \end{equation}
where $\Omega$ is the phase space. Suppose to define a probability
density function of the initial data on $\Omega$. Namely we have a function $\psi$
defined in the phase space $\Omega$, such that for all $B\subset \Omega$,
$\int_B \psi(x)dx$ denotes the probability that a randomly drawn initial datum
will belong to $B$ and $\int_{\Omega}\psi(x)dx=1$.

We are interested in determining for any $t>0$, the probability that a
solution of~(\ref{eq:ode1}) will fall in a open set $B'\subset \Omega$. Let
us call $f(x,t)$ such probability, by continuity we must have $f(x,0)=\psi(x)$
and  $\int_{\Omega}f(x,t)dx =1$ for all $t>0$. 

For any $B\subset \Omega$, $P(B)=\int_{B}f(x,t)dx$ denotes the probability
to find a point in $B$ at time $t$. We can then assume that this probability
does not change if the set $B'$ is transported by the flow of~(\ref{eq:ode1}),
$P(B)=P(A)$ where $A=\Phi^s(B)$, being $\Phi^s$ the flow at time $s$ of the
vector field. Namely
\begin{equation}
  \int_{A=\Phi^s(B)}f(y,t+s)\, dy =\int_B f(x,t)\, dx\, ,
\end{equation}
the change of coordinates $y=\Phi^s(x)$ allows to rewrite the previous
relation as follows:
\begin{equation}
\label{eq:step2}
  \int_{A=\Phi^s(B)}f(y,t+s)\, dy =\int_B f(\Phi^s(x),t+s)\det D\Phi^s(x) \,
  dx=\int_B f(x,t)\, dx\, , 
\end{equation}
being $D\Phi^s(x)$ the Jacobian of the change of variables.

The relation~(\ref{eq:step2}) should be valid for any set $B$, thus:
\begin{equation}
\label{eq:step3}
  f(x,t)= f(\Phi^s(x),t+s)\det D\Phi^s(x) \, ,
\end{equation}
for all $x\in\Omega$ and for all $t,s$. Deriving with respect to $s$ and
evaluating the derivative at $s=0$ we get the required relation (recall
$D\Phi^0(x)=identity$): 
\begin{equation}
\label{eq:therelat}
  \frac{\partial f}{\partial t}(x,t)+\nabla_x f(x,t)\cdot \vec{X}(x) +
  f(x,t)\mathit{div} \vec{X}(x)=0\, . 
\end{equation}


\begin{thebibliography}{99}
\bibitem{murray} J.D. Murray, Mathematical Biology: An introduction, Springer (1989).
\bibitem{verhulst} P.F. Verhulst, {\it Notice sur la loi que la popolation poursuit dans son accroissement},
Correspondance math\'ematique et physique, {\bf 9} 113-121 (1838)
\bibitem{Krebs} C.J. Krebs,{\it Ecology: The Experimental Analysis of Distribution and Abundance}, Harper and Row, New York (19729
\bibitem{Stogatz} S.H. Strogatz, {\it Non Linear Dynamics and Chaos}, Westview Press (2000)
\end{thebibliography}
\end{document}